\documentclass[aps, prd, twocolumn, nofootinbib,]{revtex4-2}

\usepackage{amsmath}
\usepackage{amsfonts}
\usepackage{wasysym}
\usepackage{graphicx}
\usepackage{comment}
\usepackage{tensor}
\usepackage{cancel}
\usepackage[svgnames]{xcolor}

\usepackage[colorlinks = true, linkcolor = DarkBlue, citecolor = DarkBlue, urlcolor = DarkBlue, bookmarks = false]{hyperref}

\def\filetype{pdf}
\def\path{}

\allowdisplaybreaks[1]

\begin{document}


\title{Quantum corrected Einstein-Yang-Mills black holes in semiclassical gravity}
\author{Ben Kain}
\affiliation{Department of Physics, College of the Holy Cross, Worcester, Massachusetts 01610, USA}

\begin{abstract}
\noindent
We study Einstein-Yang-Mills (EYM) black holes in semiclassical gravity by including in the field equations the expectation value of a renormalized energy-momentum tensor in the Polyakov approximation. We solve the field equations and the equations of motion self-consistently. This framework has previously been used to study vacuum solutions, i.e.~quantum corrected Schwarzschild black holes. In our system, which contains black hole hair, we find quantum corrected solutions in which the classical horizon disappears and is replaced by a wormhole structure, similarly to what is found in vacuum. We also find quantum corrected solutions in which the classical horizon disappears, but there is no wormhole structure. Our results indicate that the inclusion of black hole hair in semiclassical gravity still leads to the disappearance of classical horizons for nonextremal black holes in static spherically symmetric systems.
\end{abstract} 

\maketitle


\section{Introduction}

In semiclassical gravity, spacetime is treated classically while the source of gravity is treated using both classical and quantum methods \cite{Birrell:1982ix}. One expects that semiclassical gravity is derivable from a quantum theory of gravity and, in this sense, that semiclassical gravity is giving the first quantum corrections to the classical theory.

A central challenge in semiclassical gravity is the computation of the renormalized energy-momentum tensor. The exact renormalized energy-momentum tensor in a static spherically symmetric spacetime is complicated and contains higher order derivatives \cite{Anderson:1994hg}. The presence of these derivatives makes it difficult to see how to solve the semiclassical field equations self-consistently. As such, the exact renormalized energy-momentum tensor is typically evaluated on a fixed spacetime background.

A popular alternative is to use the Polyakov approximation and to build a spherically symmetric renormalized energy-momentum tensor in $3+1$ dimensions from the exact renormalized energy-momentum tensor computed in $1+1$ dimensions \cite{Davies:1976ei, davies, Polyakov:1981rd}. In this approximation, a much simpler analytical result can be obtained, whose behavior is expected to qualitatively agree with the behavior of the exact renormalized energy-momentum tensor and for which the semiclassical field equations can be solved self-consistently (for reviews, see \cite{Fabbri:2005mw, Arrechea:2023hmo}).

The Polyakov approximation was used to study the static spherically symmetric vacuum solution, i.e.~the quantum corrected Schwarzschild black hole, in \cite{Fabbri:2005zn}. It was found that the classical horizon is no longer present in the quantum corrected solution and is replaced by a wormhole structure. The areal radius features a minimum, forming the wormhole throat. Beyond this minimum, and hence through the wormhole, exists a curvature singularity at a finite proper distance. These results have been corroborated in various studies \cite{Berthiere:2017tms, Ho:2017joh, Ho:2017vgi, Arrechea:2019jgx}. The disappearance of the horizon and the presence of a wormhole structure were also found in the electrovacuum, i.e.~the quantum corrected Reissner-Nordstr\"om black hole \cite{Arrechea:2021ldl}.  Related studies of stellar configurations and alternatives to the Polyakov approximation include \cite{Carballo-Rubio:2017tlh, Arrechea:2021pvg, Beltran-Palau:2022nec, Arrechea:2022dvy, Arrechea:2023oax}.

The results in \cite{Fabbri:2005zn, Berthiere:2017tms, Ho:2017joh, Ho:2017vgi, Arrechea:2019jgx, Arrechea:2021ldl} indicate that quantum corrections can have dramatic consequences for classical black holes. The fact that quantum corrections can remove the horizon from a classical black hole is fascinating. We are interested in the extent to which this can occur. Schwarzschild and Reissner-Nordstr\"om are the canonical spherically symmetric black hole solutions, but they are not the only spherically symmetric black hole solutions available. In this work, we study how black hole hair might affect these results. More specifically, if we begin with a classical hairy black hole solution, how is the solution modified by the inclusion of a renormalized energy-momentum tensor in the framework of semiclassical gravity? Will the classical horizon disappear? If so, is there a wormhole?

We study Einstein-Yang-Mills (EYM) black holes \cite{Volkov:1989fi, Bizon:1990sr, Kuenzle:1990is}. These are classical spherically symmetric static black hole solutions with $SU(2)$ Yang-Mills matter. In fact, these were the first hairy black holes discovered \cite{Volkov:2016ehx}. Using the semiclassical gravity framework, we include a renormalized energy-momentum tensor in the Polyakov approximation. We numerically solve the equations self-consistently to find quantum corrected EYM black holes.

Classical EYM black holes are identified in part by the classical horizon radius. We find that, for sufficiently large values of the horizon radius, the quantum corrected solutions are similar to vacuum solutions, in that the classical horizon disappears and is replaced by a wormhole structure. On the other hand, for smaller values of the classical horizon radius, the classical horizon still disappears, but there is no wormhole. Our results show that semiclassical corrections causing the disappearance of classical horizons is a robust phenomenon, occurring in a broad range of classical static black hole solutions. 

In Sec.~\ref{sec:YM}, we describe Yang-Mills (YM) matter and in Sec.~\ref{sec:classical sols}, we review classical EYM black holes. In Sec.~\ref{sec:quantum}, we review the renormalized energy-momentum tensor in the Polyakov approximation, construct the field equations, study boundary conditions, and explain how we numerically compute our solutions. We present numerical results in Sec.~\ref{sec:results} and derive approximate analytical formulas in Sec.~\ref{sec:analytical}. We conclude in Sec.~\ref{sec:conclusion}. We use units such that $G=c=\hbar=1$ throughout.


\section{Yang-Mills matter}
\label{sec:YM}

We will be studying static spherically symmetric solutions where the matter content is a pure Yang-Mills theory with gauge group $SU(2)$. The gauge field is $A_\mu^a$, where $\mu$ is the spacetime index and $a=1,2,3$ is the gauge index. We will use a couple of different forms for the static spherically symmetric metric. To make moving between these forms as simple as possible, we will present equations in this section using the general form
\begin{equation} \label{general metric}
ds^2 = -\alpha^2(x) dt^2 + a^2(x) dx^2 + r^2(x) d\Omega^2,
\end{equation}
where $x$ is the radial coordinate, $\alpha(x)$, $a(x)$, and $r(x)$ are the metric fields, and $d\Omega^2 \equiv  d\theta^2 + \sin^2\theta \, d\phi^2$.

The pure Yang-Mills Lagrangian is
\begin{equation} \label{Lagrangian}
\mathcal{L}_\text{YM} = - \frac{1}{4g^2} F_{\mu\nu}^a F^{\mu\nu}_a,
\end{equation}
where $g$ is the coupling constant,
\begin{equation} \label{field strength}
F_{\mu\nu}^a = \nabla_\mu A_\nu^a - \nabla_\nu A_\mu^a + \epsilon_{abc} A_\mu^b A_\nu^c
\end{equation}
is the field strength, and $\epsilon_{abc}$ is the completely antisymmetric symbol with $\epsilon_{123} = 1$. Gauge indices can be placed up or down and repeated gauge indices are summed over. For $SU(2)$, the group algebra is $[T_a, T_b] = i \epsilon_{abc} T_c$, where $T_a$ are the generators of the group. It is convenient to define 
\begin{equation}
A_\mu \equiv T^a A_\mu^a, \qquad
F_{\mu\nu} \equiv T^a F^a_{\mu\nu}.
\end{equation}
Gauge transformations are then given by
\begin{equation}
\begin{split}
A_\mu &\rightarrow A_\mu' = U A_\mu U^{-1} - i\left(\nabla_\mu U \right) U^{-1}
\\
F_{\mu\nu} &\rightarrow F_{\mu\nu}' = U F_{\mu\nu} U^{-1},
\end{split}
\end{equation}
where $U = e^{-i\Lambda}$ and $\Lambda = \Lambda^a T^a$ with $\Lambda^a$ the gauge functions.

In spherical symmetry, the gauge field is constrained to take the form \cite{Witten:1976ck, Bartnik:1988am, Volkov:1998cc}
\begin{equation} \label{gauge field}
\begin{split}
A_t &= T^3 u
\\
A_x &= T^3 v
\\
A_\theta &= T^1 w_1 + T^2 w_2
\\
A_\phi &= \left(-T^1 w_1 + T^2 w_2 + T^3 \cot\theta \right)\sin\theta,
\end{split}
\end{equation}
where $u(x)$, $v(x)$, $w_1(x)$, and $w_2(x)$ are the four gauge fields that parametrize $A_\mu$. The gauge field in (\ref{gauge field}) is written in a gauge where the $T^a$ take their standard constant form. Inserting the gauge field into (\ref{field strength}), we find for the nonvanishing components of the field strength
\begin{align} 
F_{tx} &= -T^3  \partial_x u 
\notag \\
 F_{t\theta} &=
-T^1 u w_1 + T^2 u w_2
\notag \\
F_{t\phi} &=
- (T^1 u w_2 + T^2 u w_1) \sin\theta 
\notag \\
F_{x\theta} &= 
T^1 \left(\partial_x w_2 - v w_1 \right)
+ T^2 \left(\partial_x w_1 + v w_2 \right) 
\notag \\
F_{x\phi} &=
- \bigl[ T^1 \left(\partial_x w_1 + v w_2 \right) - T^2 \left(\partial_x w_2 - v w_1 \right) \big] \sin\theta 
\notag \\
F_{\theta\phi} &= 
-T^3 \left(1 - w_1^2 - w_2^2 \right)  \sin\theta.
\label{field strength comps}
\end{align}
Inserting the field strength into (\ref{Lagrangian}), we find for the Lagrangian
\begin{equation} \label{Lagrangian 2}
\begin{split}
g^2 \mathcal{L}_\text{YM}
&= \frac{(\partial_x u)^2}{2 \alpha^2 a^2} 
- \frac{(1-w_1^2 - w_2^2)^2}{2 r^4}
+ \frac{u^2(w_1^2 + w_2^2)}{\alpha^2 r^2}
\\
&\qquad
- \frac{(\partial_x w_2 - v w_1)^2 + (\partial_x w_1 + v w_2)^2}{a^2 r^2}.
\end{split}
\end{equation}

Spherical symmetry breaks $SU(2)$ down to $U(1)$ \cite{Witten:1976ck}. The Lagrangian in (\ref{Lagrangian 2}) has the residual $U(1)$ invariance 
\begin{equation} \label{U(1) symmetry}
v \rightarrow v' = v - \partial_x \lambda, \qquad
w \rightarrow w' = w e^{-i\lambda},
\end{equation}
and $u$ unchanged, where $w = w_1 + i w_2$ and $\lambda$ is the gauge parameter.  We can use this $U(1)$ invariance to simplify the theory. It is possible to choose a gauge such that $v = w_2 = 0$. We will make an additional simplification, which is called the magnetic ansatz. We set $u = 0$. This is a constraint that we impose on the theory, and not a gauge choice, which leads to a consistent set of equations. 

With these simplifications, the Lagrangian becomes
\begin{equation} \label{simplified Lagrangian}
\mathcal{L}_\text{YM}
= - \frac{(\partial_x w)^2}{g^2 a^2 r^2} - \frac{(1 - w^2)^2}{2 g^2 r^4},
\end{equation}
where $w\equiv w_1$ is the sole matter field. Having established the matter Lagrangian, we can minimally couple it to gravity and form the Lagrangian $\mathcal{L} = \sqrt{-\det(g_{\mu\nu})} \mathcal{L}_\text{YM}$, where $\det(g_{\mu\nu}) = \alpha^2 a^2 r^4 \sin^2\theta$ is the determinant of the metric. From this Lagrangian, the equation of motion is 
\begin{equation} \label{general eom}
0 = \partial_x \left( \frac{\alpha}{a} \partial_x w \right) + \frac{\alpha a}{r^2} w (1-w^2).
\end{equation}

The energy-momentum tensor, as derived from the Lagrangian in (\ref{Lagrangian}), is given by
\begin{equation}
T_{\mu\nu} = 
\frac{1}{g^2}g^{\sigma\lambda} F^a_{\mu\sigma} F^a_{\nu\lambda}
+ g_{\mu\nu} \mathcal{L}_\text{YM}.
\end{equation}
We compute the first term using the field strength components in (\ref{field strength comps}) after setting the relevant fields to zero. We compute the second term using the simplified Lagrangian in (\ref{simplified Lagrangian}). The nonvanishing components of the energy-momentum tensor are
\begin{equation} \label{general T}
\begin{split}
T\indices{^t_t} 
&= 
- \frac{(1-w^2)^2}{2 g^2 r^4}
- \frac{(\partial_x w)^2}{g^2 a^2 r^2}
\\
T\indices{^x_x}
&=
- \frac{(1-w^2)^2}{2 g^2 r^4}
+ \frac{(\partial_x w)^2}{g^2 a^2 r^2}
\\
T\indices{^\theta_\theta} 
&=
+ \frac{(1-w^2)^2}{2 g^2 r^4},
\end{split}
\end{equation}
along with $T\indices{^\phi_\phi} = T\indices{^\theta_\theta}$.


\section{Classical Einstein-Yang-Mills black holes}
\label{sec:classical sols}

In this section, we review classical solutions of EYM black holes \cite{Volkov:1989fi, Bizon:1990sr, Kuenzle:1990is, Volkov:1998cc}. In the next section, we shall find that classical solutions at large radial coordinate are still valid solutions when vacuum energy is included. As such, the classical solutions we review in this section will be used for the outer boundary values when solving for quantum corrected solutions.

In computing classical solutions, it is convenient to write the static spherically symmetric metric in the form
\begin{equation} \label{classical metric}
ds^2 = - \sigma^2(r) N(r) dt^2 + \frac{1}{N(r)} dr^2 + r^2 d\Omega^2,
\end{equation}
where
\begin{equation} \label{N def}
N(r) = 1- \frac{2m(r)}{r}.
\end{equation}
This form allows the relevant boundary conditions to be straightforwardly implemented, as we will see below. The metric in (\ref{classical metric}) uses the areal radius, $r$, as the radial coordinate and is parametrized in terms of the metric functions $\sigma(r)$ and $m(r)$. $m(r)$ gives the total mass inside a radius $r$. The ADM mass, $M$, is given by $m(r)$ in the large $r$ limit.

The $tt$ and $rr$ components of the Einstein field equations lead to
\begin{equation} \label{classical field equations}
\begin{split}
\partial_r\sigma&= 4\pi \frac{r \sigma}{N} \left( T\indices{^r_r} - T\indices{^t_t} \right)
\\
\partial_r m &= -4\pi r^2 T\indices{^t_t}.
\end{split}
\end{equation}
Comparing the general metric in (\ref{general metric}) to the metric we are using in this section in (\ref{classical metric}), we find
\begin{equation}
x\rightarrow r,
\qquad
\alpha \rightarrow \sigma \sqrt{N},
\qquad
a \rightarrow \frac{1}{\sqrt{N}}.
\end{equation}
With these replacements, the equation of motion in (\ref{general eom}) becomes
\begin{equation}  \label{classical eom}
\partial_r^2 w = - \left[  4\pi r \left( T\indices{^r_r} + T\indices{^t_t} \right) 
+ \frac{2m}{r^2}  \right] \frac{\partial_r w}{N}
- \frac{w(1- w^2)}{N r^2},
\end{equation}
where we used (\ref{classical field equations}) to write the equation of motion in this form, and the energy-momentum tensor in (\ref{general T}) becomes
\begin{equation} \label{classical T}
\begin{split}
T\indices{^t_t} &= - \frac{(1-w^2)^2}{2g^2 r^4} - \frac{N (\partial_r w)^2}{g^2 r^2}
\\
T\indices{^r_r} &= - \frac{(1-w^2)^2}{2g^2 r^4} + \frac{N (\partial_r w)^2}{g^2 r^2}
\\
T\indices{^\theta_\theta} 
&= + \frac{(1-w^2)^2}{2 g^2 r^4}.
\end{split}
\end{equation}

EYM black holes are solutions to the system of equations in (\ref{classical field equations}) and (\ref{classical eom}). 
$\sigma$ only shows up in the top equation in (\ref{classical field equations}), making it easy to see that for a given solution, we can scale $\sigma$ by a constant and it is still a solution. We will make use of this symmetry below. It is convenient to introduce the scaled quantities
\begin{equation} \label{classical scaling}
\bar{r} \equiv g r,
\qquad
\bar{m} \equiv g m,
\end{equation}  
for which $N = 1 - 2\bar{m}/\bar{r}$. When equations are written in terms of $\bar{r}$ and $\bar{m}$, the coupling constant $g$ disappears and does not have to be specified.

To solve the scaled version of the system of equations in (\ref{classical field equations}) and (\ref{classical eom}), we need boundary conditions. We begin with inner boundary conditions. If we expand quantities around $r = 0$, we will find boundary conditions that lead to the regular Bartnik-McKinnon solutions \cite{Bartnik:1988am}. To find black hole solutions, we choose a horizon radius, $\bar{r}_h$, and require that
\begin{equation}
N(\bar{r}_h) = 0.
\end{equation} 
This boundary condition sets the metric component $g_{tt}$ to zero at the horizon, as can be seen from (\ref{classical metric}), and introduces a coordinate singularity for $g_{rr}$, as is appropriate since the metric in (\ref{classical metric}) is parametrized in terms of the Schwarzschild radial coordinate.

We expand quantities about $\bar{r} = \bar{r}_h$,
\begin{equation} \label{iBC}
\begin{split}
\sigma &= \sigma_h + \sigma_1(\bar{r} - \bar{r}_h) + \cdots
\\
w &= w_h + w_1(\bar{r} - \bar{r}_h) + \cdots,
\\
N &= N_1(\bar{r} - \bar{r}_h) + \cdots. 
\end{split}
\end{equation}
Plugging these expansions into the system of equations, we find the solutions
\begin{equation} \label{iBC sols}
\begin{split}
\sigma_1 &= \sigma_h \frac{8\pi w_1^2}{\bar{r}_h}
\\
w_1 &= \frac{\bar{r}_h w_h (w_h^2  - 1)}{\bar{r}_h^2 - 4\pi (w_h^2 - 1)^2}
\\
N_1 &= \frac{1}{\bar{r}_h} \left[ 1 - \frac{4\pi(w_h^2 - 1)^2}{\bar{r}_h^2} \right].
\\
\end{split}
\end{equation}
From (\ref{N def}),
\begin{equation} \label{m iBC}
\bar{m} = \frac{\bar{r}_h}{2} + \frac{1}{2} (1 - N_1 \bar{r}_h)(\bar{r} - \bar{r}_h) + \cdots.
\end{equation}
The total mass contained inside the horizon of EYM black holes is equal to $\bar{r}_h/2$. Of course, there are additional contributions to the mass outside the horizon from the YM matter. Equations (\ref{iBC})--(\ref{m iBC}), which are parametrized in terms of the constants $\bar{r}_h$, $\sigma_h$, and $w_h$, are the inner boundary conditions we will use. 

To compute solutions we choose a value for $\bar{r}_h$ and set $\sigma_h = 1$. We use the shooting method, integrating outward from $\bar{r}_h$ and tuning the value of $w_h$ until the outer boundary conditions are satisfied. We will present the derivation of the outer boundary conditions in the next section. For now, we simply state that the outer boundary condition is $w \rightarrow \pm 1$ as $\bar{r} \rightarrow \infty$. Once a solution is found, we can use the above mentioned symmetry to scale $\sigma$ such that $\sigma \rightarrow 1$ as $\bar{r} \rightarrow \infty$. This is the correct asymptotic value for $\sigma$ such that the metric takes the standard asymptotically flat form. We can also compute the ADM mass as 
\begin{equation} \label{ADM M}
\overline{M} = \bar{m}(\bar{r} \rightarrow \infty),
\end{equation}
where $\overline{M} \equiv g M$.

\begin{figure} 
\includegraphics[width=2.8in]{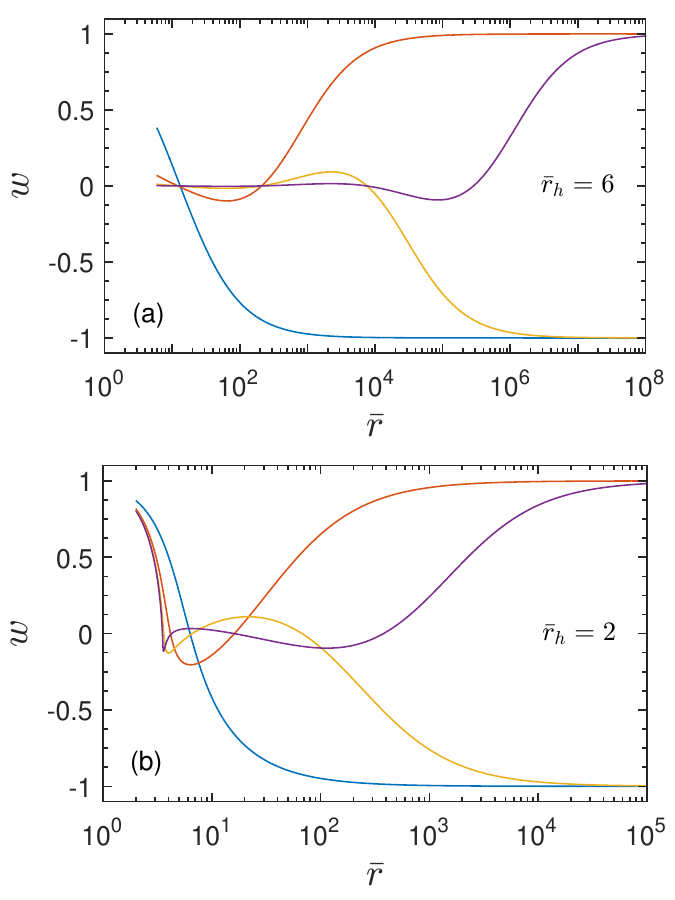} 
\caption{Yang-Mills matter field $w$ as a function of the radial coordinate $\bar{r}$ for classical EYM black holes for (a) $\bar{r}_h = 6$ and (b) $\bar{r}_h = 2$, where $\bar{r}_h$ is the horizon radius. In each plot, solutions are shown for $n=1$ to 4 radial nodes.}
\label{fig:classical solutions}
\end{figure}

We show solutions for $w$ in Fig.~\ref{fig:classical solutions}(a) for $\bar{r}_h = 6$ and in Fig.~\ref{fig:classical solutions}(b) for $\bar{r}_h = 2$. Note that all solutions have at least one radial node. Nodeless solutions, for which $w=1$ for all $\bar{r} \geq \bar{r}_h$, have a vanishing energy-momentum tensor and are the vacuum Schwarzschild solution. Because of this, EYM solutions cannot be continuously connected to the Schwarzschild solution. EYM black holes are uniquely identified by the number of nodes, $n$, and the horizon radius, $\bar{r}_h$. In the limit $\bar{r}_h\rightarrow 0$, the solutions become the regular Bartnik-McKinnon solutions \cite{Bartnik:1988am}.

\begin{figure}
\includegraphics[width=3.4in]{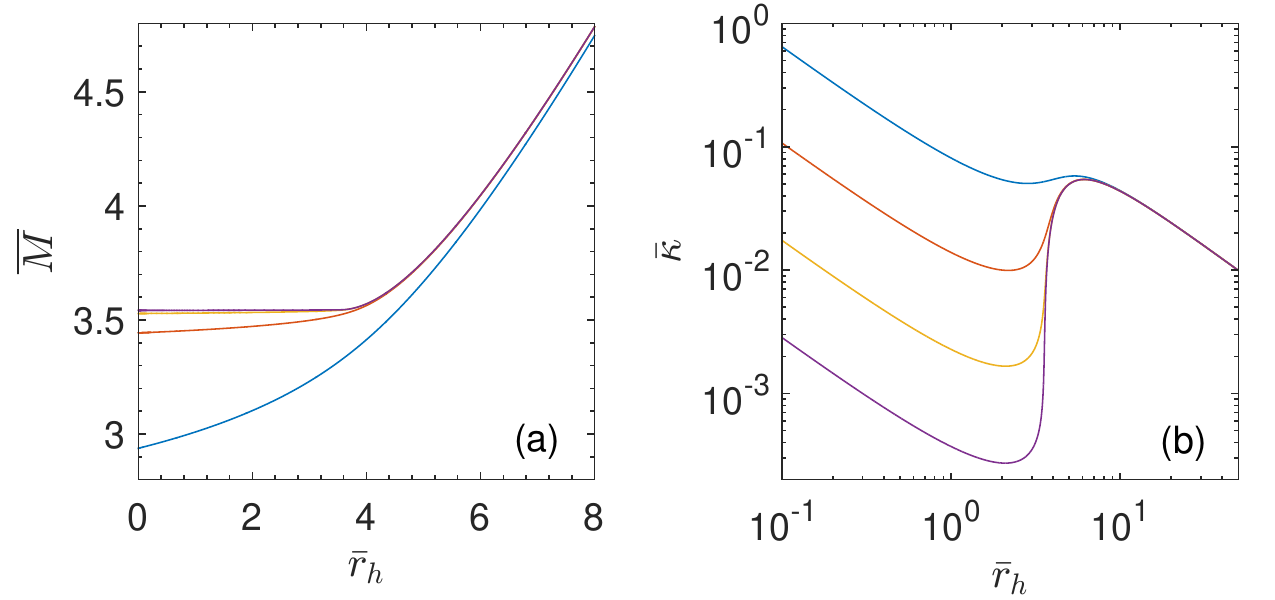} 
\caption{(a) ADM mass $\overline{M}$ as a function of horizon radius $\bar{r}_h$ for classical EYM black holes with, from bottom to top, $n=1$ to 4 nodes. (b) Surface gravity $\bar{\kappa}$ with, from top to bottom, $n = 1$ to 4 nodes.}
\label{fig:classical Mvr}
\end{figure}

In Fig.~\ref{fig:classical Mvr}(a), we plot the ADM mass, $\overline{M}$, as a function of the horizon radius, $\bar{r}_h$, for $n=1$ to 4 nodes. For a given $\bar{r}_h$, we see that $\overline{M}$ increases for increasing $n$. For this reason, solutions with a single node are referred to as fundamental solutions and solutions with more than one node are referred to as excited solutions. For larger values of $\bar{r}_h$, it is clear that the ADM mass is effectively given by the large $n$ value. Figure \ref{fig:classical solutions}(a) indicates that in the limit $n\rightarrow \infty$, we have $w\rightarrow 0$ for all $\bar{r} \geq \bar{r}_h$. Setting $w = 0$ in the energy-momentum tensor in (\ref{classical T}), we can compute the ADM mass analytically in this case using the bottom equation in (\ref{classical field equations}),
\begin{equation} \label{M large n rh}
M_{n=\infty} = \frac{\bar{r}_h}{2} + \frac{2\pi}{\bar{r}_h}.
\end{equation}

Figure \ref{fig:classical Mvr}(a) and Eq.~(\ref{M large n rh}) suggest that for larger horizon radius, the YM matter has a minimal effect on the spacetime. On the other hand, as the horizon radius decreases, the effect of the YM matter becomes more pronounced. For smaller $\bar{r}_h$, we can see from Fig.~\ref{fig:classical Mvr}(a) that, for $n > 1$, the ADM mass is well approximated by the large $n$ value for $\bar{r}_h = 0$. This can be shown to be \cite{Bartnik:1988am}
\begin{equation}
\overline{M}_{n=\infty} = \sqrt{4\pi}.
\end{equation}

We end this section with a computation of the surface gravity for EYM black holes. For the general static spherically symmetric metric in (\ref{general metric}), the surface gravity is given by $\kappa = (\partial_x \alpha)/a$, where the right-hand side is to be evaluated at the horizon. We assume the conventional normalization, which for the metric in (\ref{general metric}) is $\alpha \rightarrow 1$ as $x \rightarrow \infty$. For EYM black holes, as parametrized in this section, the surface gravity is given by
\begin{equation}
\bar{\kappa} = \frac{1}{2} \sigma_h N_1,
\end{equation}
where $\bar{\kappa} \equiv \kappa/g$. In Fig.~\ref{fig:classical Mvr}(b), we plot the surface gravity as a function of the horizon radius for $n=1$ to 4 nodes. We can see that EYM black holes have nonzero surface gravity, and hence are nonextremal, except in the limit $n\rightarrow \infty$ \cite{Volkov:1998cc}. 


\section{Quantum corrected Einstein-Yang-Mills black holes}
\label{sec:quantum}

We turn now to the main topic of this work. We would like to self-consistently solve the semiclassical Einstein field equations \cite{Birrell:1982ix, Fabbri:2005mw, Arrechea:2023hmo},
\begin{equation} \label{field equations}
G_{\mu\nu} = 8\pi \left( T_{\mu\nu} + \langle \widehat{T}_{\mu\nu} \rangle \right),
\end{equation}
where $G_{\mu\nu}$ is the Einstein tensor and $T_{\mu\nu}$ is a classical energy-momentum tensor describing the matter in the system. $\langle \widehat{T}_{\mu\nu} \rangle$ is the quantum contribution to the matter sector, which we take to be the expectation value of a renormalized energy-momentum tensor operator in the vacuum state in the Polyakov approximation. We present its form in Sec.~\ref{sec:RSET}.

To proceed, we must choose a form for the static spherically symmetric metric. At this point, it is unclear how the vacuum polarization included in the semiclassical field equations will affect the spacetime. For example, will the horizon present in the classical solution survive or will it disappear and be replaced by a wormhole? It is convenient to choose a form for the metric that can accommodate numerical solutions containing horizons or horizonless wormholes. We have found that 
\begin{equation} \label{metric}
ds^2 = e^{2\nu(x)} (-dt^2 + dx^2) + r^2(x) d\Omega^2
\end{equation}
works well, where $\nu(x)$ and $r(x)$ parametrize the metric. We shall refer to the radial coordinate, $x$, as the tortoise coordinate, since this is its name when the metric in (\ref{metric}) describes a Schwarzschild spacetime.

For the metric in (\ref{metric}), the nonzero components of the Einstein tensor are
\begin{equation} \label{Einstein tensor}
\begin{split}
G_{tt} &= \frac{1}{r^2} \left(  2r r'\nu' -r^{\prime\, 2} -2rr'' + e^{2\nu}\right)
\\
G_{xx} &= \frac{1}{r^2} \left( 2rr'\nu' + r^{\prime\,2} - e^{2\nu} \right)
\\
G_{\theta\theta} &= r e^{-2\nu} \left( r'' + r \nu'' \right),
\end{split}
\end{equation}
along with $G_{\phi\phi} = G_{\theta\theta} \sin^2\theta$, where, from this point forward, a prime denotes an $x$ derivative.


\subsection{Renormalized energy-momentum tensor}
\label{sec:RSET}

As mentioned, we construct the quantum contribution to the semiclassical field equations, $\langle \widehat{T}_{\mu\nu} \rangle$, using the Polyakov approximation. In the Polyakov approximation, we build the renormalized energy-momentum tensor in $3+1$ dimensions from the exact renormalized energy-momentum tensor in $1+1$ dimensions. 

To keep things simple, we shall compute the renormalized energy-momentum tensor describing vacuum polarization in $1+1$ dimensions for a single massless scalar field, as is commonly done \cite{Davies:1976ei, davies}. A convenient formula for this quantity was presented in \cite{Barcelo:2011bb}. For the two-dimensional metric $ds^2 = g_{ab} dx^a dx^b$, where $a,b = t,x$, we have
\begin{equation} \label{1+1}
\langle \widehat{T}_{ab}\rangle^{\text{(2D)}}
= \frac{R^{\text{(2D)}}}{48\pi} g_{ab} + \frac{1}{48\pi} \left( \mathcal{A}_{ab} - \frac{1}{2} g_{ab} \mathcal{A}\indices{^c_c} \right),
\end{equation}
where $R^{\text{(2D)}}$ is the two-dimensional Ricci scalar and where
\begin{equation}
\mathcal{A}_{ab} = \frac{4}{|\xi|} \nabla_a \nabla_b |\xi|,
\qquad
|\xi| = \sqrt{-\xi_a \xi^a}.
\end{equation}
The quantity $\xi^a$ parametrizes the vacuum state, with respect to which the expectation value is calculated. Since we are assuming a static and asymptotically flat spacetime, the appropriate choice is the Boulware vacuum, for which $\xi^a = (1,0)$.

In spherical symmetry, the nonangular dimensions are an effective $1+1$ dimensional spacetime. In the Polyakov approximation, we build the renormalized energy-momentum tensor in spherically symmetric $3+1$ dimensions using the $1+1$ result for the nonangular components,
\begin{equation}
\langle \widehat{T}_{ab}\rangle = F(r) \langle \widehat{T}_{ab}\rangle^{\text{(2D)}},
\end{equation}
where $F(r)$ is a multiplicative factor that depends on the areal radius, $r$. We will give the standard form for $F(r)$ in a moment. We require $3+1$-dimensional conservation, $\nabla_\mu \langle \widehat{T}\indices{^\mu_\nu}\rangle = 0$. In general this can be accomplished by introducing angular components, which can be solved for algebraically from the conservation equation.

For the metric in (\ref{metric}), the corresponding two-dimensional metric is $ds^2 = e^{2\nu}(-dt^2 + dx^2)$. The renormalized energy-momentum tensor in the Polyakov approximation is then
\begin{equation} \label{RSET}
\begin{split}
\langle \widehat{T}_{tt} \rangle 
&= - \frac{F}{24\pi} \left(\nu^{\prime\,2} - 2\nu'' \right)
\\
\langle \widehat{T}_{xx} \rangle 
&= -\frac{F}{24\pi} \nu^{\prime\,2}
\\
\langle \widehat{T}_{\theta\theta} \rangle 
&= - \frac{1}{24\pi} r^2 e^{-2\nu} \nu^{\prime\,2} \left(F + \frac{r F'}{2r'} \right),
\end{split}
\end{equation}
along with $\langle \widehat{T}_{\phi\phi} \rangle  = \langle \widehat{T}_{\theta\theta} \rangle  \sin^2\theta$. The standard choice for $F$ is
\begin{equation} \label{upscale}
F = \frac{1}{4\pi r^2}.
\end{equation}
For this choice, it immediately follows that $F + rF'/2r' = 0$ and that $\langle \widehat{T}_{\theta\theta} \rangle = 0$. Indeed, this choice for $F$ is often explained as ensuring conservation without having to introduce angular components. The disadvantage of this choice is that it leads to a divergence in the renormalized energy-momentum tensor as $r \rightarrow 0$. This can be avoided by using an $F$ which regulates the divergence \cite{Parentani:1994ij, Ayal:1997ab, Arrechea:2019jgx}. We will not consider such a choice here, and will use (\ref{upscale}), but we must be mindful of the divergence if solutions approach $r=0$. We will continue to write formulas in terms of $F$ for completeness.

The Polyakov approximation, from the perspective of $3+1$ dimensions, can be viewed as making two approximations to the field equations for the massless scalar field. The approximations are, first, ignoring all but the $s$-wave component in a decomposition of the field in spherical harmonics and, second, dropping the $s$-wave component of the potential in the equations of motion. A particular value of the Polyakov approximation is the simple form for the renormalized energy-momentum tensor in (\ref{RSET}), a form that allows one to self-consistently solve the field equations. Further, the behavior of (\ref{RSET}) is expected to capture qualitatively the behavior of the exact renormalized energy-momentum tensor, as long as we stay away from $r = 0$.


\subsection{Equations:~$\nu(x)$, $r(x)$, and $w(x)$}
\label{sec:equations}

We are now in a place to derive the equations whose solutions describe quantum corrected EYM black holes. Comparing the general metric in (\ref{general metric}) to the metric we are using in this section in (\ref{metric}), we find
\begin{equation}
\alpha \rightarrow e^\nu,
\qquad
a \rightarrow e^\nu.
\end{equation}
With these replacements, the equation of motion in (\ref{general eom}) becomes
\begin{equation} \label{eom}
w'' = -\frac{e^{2\nu}}{r^2} w (1-w^2)
\end{equation}
and the energy-momentum tensor in (\ref{general T}) becomes
\begin{equation} \label{T}
\begin{split}
T\indices{^t_t} &= -\frac{(1 - w^2)^2}{2g^2 r^4} 
- \frac{w^{\prime\,2}}{g^2 r^2 e^{ 2\nu}}
\\
T\indices{^r_r} &= -\frac{(1 - w^2)^2}{2g^2 r^4} 
+ \frac{w^{\prime\,2}}{g^2 r^2 e^{ 2\nu}}
\\
T\indices{^\theta_\theta} &= +\frac{(1 - w^2)^2}{2g^2 r^4} .
\end{split}
\end{equation}

The semiclassical field equations in (\ref{field equations}) require a little more work. We are interested in deriving ordinary differential equations (ODEs) for the metric fields $\nu$ and $r$. The Einstein tensor is given in (\ref{Einstein tensor}) and the renormalized energy-momentum tensor is given in (\ref{RSET}). These are to be plugged into the field equations along with (\ref{T}). Adding together the $tt$ and $rr$ components of the field equations leads to
\begin{equation}
r'' - 2r'\nu' = 4\pi r e^{2\nu} \left( T\indices{^t_t} - T\indices{^r_r} \right)
+ \frac{F r}{3} \left(\nu^{\prime\,2} - \nu'' \right).
\end{equation}
The $\theta\theta$ component is
\begin{equation}
r'' + r \nu''
=  8\pi r e^{2\nu} T\indices{^\theta_\theta} - \frac{1}{3} r  \nu^{\prime\,2} \left(F + \frac{r F'}{2r'} \right) .
\end{equation}
Both of these equations contain $r''$ and $\nu''$. Combining them in such a way as to eliminate one or the other, leads to
\begin{equation} \label{nu r ODEs}
\begin{split}
\nu'' \left(1 - \frac{F}{3} \right)
&= - 4\pi e^{2\nu} \left( T\indices{^t_t} - T\indices{^r_r} - 2T\indices{^\theta_\theta}\right)
\\
&\qquad
- \frac{2 r' \nu'}{r}
- \nu^{\prime\,2} \left[\frac{F}{3} + \frac{1}{3}  \left(F + \frac{r F'}{2r'} \right) \right]
\\
r'' \left(1 - \frac{F}{3} \right)
&=  4\pi r e^{2\nu} \left( T\indices{^t_t} - T\indices{^r_r} - \frac{2F}{3} T\indices{^\theta_\theta}\right)
\\
&\qquad
+ 2 r' \nu'
+ \frac{F r}{3} \nu^{\prime\,2} \left[1 + \frac{1}{3}  \left(F + \frac{r F'}{2r'} \right) \right] .
\end{split}
\end{equation}
These are the equations we are after. Quantum corrected solutions are solutions to (\ref{eom}) and (\ref{nu r ODEs}). We note that with the inclusion of the vacuum polarization we can no longer remove the coupling constant $g$ through a redefinition of fields, as we did with the classical equations.

There is an additional equation that is useful. It follows directly from the $rr$ component of the field equations,
\begin{equation} \label{rr field equation}
2rr'\nu' + r^{\prime\,2} - e^{2\nu}
= 8\pi r^2 e^{2\nu} T\indices{^r_r} -\frac{F}{3} r^2 \nu^{\prime \, 2},
\end{equation}
which is quadratic in $\nu'$. Solving for $\nu'$, we find
\begin{equation} \label{nu' pm}
\nu' 
= - \frac{3 r'}{F r} \left[
1 \pm \sqrt{
1 + \frac{F }{3 r^{\prime\,2}} \left( 8\pi r^2 e^{2\nu} T\indices{^r_r} - r^{\prime \, 2} + e^{2\nu} \right)
}\right].
\end{equation}
The classical limit is the limit in which the renormalized energy-momentum tensor is sent to zero, which is obtained in the above equation by $F\rightarrow 0$. If we use the lower sign in (\ref{nu' pm}), then this limit leads to the standard classical result. With the lower sign, then, the renormalized energy-momentum tensor can be thought of as a quantum perturbation. On the other hand, with the upper sign, the classical limit does not lead to the classical equation. With the upper sign, then, we have a nonperturbative correction to the classical result. 

We shall find that both signs in (\ref{nu' pm}) show up in the quantum corrected solutions. As we explain in Sec.~\ref{sec:obc}, we use classical solutions as outer boundary conditions. As a consequence, the quantum corrected solution for large $x$ is given by the perturbative result, i.e.~(\ref{nu' pm}) with the lower sign. As we move inward to smaller $x$, there is a value of $x$ where some quantum corrected solutions switch continuously from the perturbative result to the nonperturbative result.


\subsection{Equations:~$\nu(r)$ and $w(r)$}
\label{sec:nu(r) w(r)}

In the previous subsection, we presented the system of equations where the tortoise coordinate, $x$, is the independent variable. We find $x$ to be a particularly good coordinate to use numerically and, as such, we prefer to use the equations presented in the previous subsection for computing numerical solutions. With that being said, it is possible to decouple the equations in (\ref{nu r ODEs}), finding a single equation for $\nu(r)$. Since we can also find an equation of motion for $w(r)$, we can construct a system of equations where $r$ is the independent variable. We will derive these equations in this subsection and we will make use of them when we study outer boundary conditions and analytical approximations.

To simplify notation, we use an overdot to indicate an $r$ derivative, so that $\dot{\nu} = \partial_r \nu$ and $\dot{w} = \partial_r w$. $x$ derivatives may be converted to $r$ derivatives with
\begin{equation} \label{deriv convert}
\nu' = r' \dot{\nu},
\qquad
\nu''= r'' \dot{\nu} + r^{\prime\,2} \ddot{\nu},
\end{equation}
and similarly for derivatives of $w$. Using these in the second equation in (\ref{nu r ODEs}), we find
\begin{equation} \label{r'' convert}
\begin{split}
r'' \left(1 - \frac{F}{3} \right)
&=  4\pi r e^{2\nu} \left( T\indices{^t_t} - T\indices{^r_r} - \frac{2 F}{3} T\indices{^\theta_\theta}\right)
+ 2 r^{\prime\,2} \dot{\nu}
\\
&\qquad
+ \frac{ F r}{3} r^{\prime\,2} \dot{\nu}^2 \left[1 + \frac{1}{3}  \left(F + \frac{r F'}{2r'} \right) \right].
\end{split}
\end{equation}
Making a similar change to the first equation in (\ref{nu r ODEs}) and then combining the results so as to eliminate $r''$, we arrive at
\begin{align} 
0 &= \ddot{\nu}  \left(1  - \frac{ F }{3} \right)
+ \dot{\nu}^3 r \frac{ F}{3} \left( 1 +  \frac{2F + rF'/r'}{6}  \right)
\notag \\
&\quad
+ \dot{\nu}^2 \left( 2 + \frac{ F }{3}
+  \frac{2F + rF'/r'}{6}  \right)
+ \frac{2\dot{\nu}}{r}
\label{rp nu ODE}
\\
&\quad 
+  4\pi \frac{ e^{2\nu}}{r^{\prime\,2}} \biggl[ (1 + r\dot{\nu})(T\indices{^t_t} - T\indices{^r_r}) 
- 2 \left( 1 + \frac{ F}{3} r \dot{\nu} \right) T\indices{^\theta_\theta} \biggr].
\notag
\end{align}
Since $F=F(r)$, we have that $F'/r' = \dot{F}$ is independent of $r'$. In vacuum, where the classical energy-momentum tensor is zero, $r'$ would be absent in the equation above and we would have succeeded in decoupling the equations in (\ref{nu r ODEs}). Since the classical energy-momentum tensors are nonzero, we must find an expression for $r'$.

We can find an expression for $r'$ from the $rr$ component of the field equations in (\ref{rr field equation}). Converting the derivatives using (\ref{deriv convert}) gives
\begin{equation} \label{r prime squared eq}
2r r^{\prime\,2} \dot{\nu} + r^{\prime\,2} - e^{2\nu} 
= 8\pi r^2 e^{2\nu} T\indices{^r_r} -\frac{F}{3} r^2 r^{\prime\,2} \dot{\nu}^2.
\end{equation}
The energy-momentum tensor in (\ref{T}) depends on $r'$ since, upon converting the $x$ derivatives in $w'$, we have
\begin{equation}
\begin{split}
T\indices{^t_t} &= -\frac{(1 - w^2)^2}{2g^2 r^4} 
- \frac{r^{\prime\,2} \dot{w}^2}{g^2 r^2 e^{ 2\nu}}
\\
T\indices{^r_r} 
&= -\frac{(1 - w^2)^2}{2g^2 r^4} + \frac{r^{\prime\,2} \dot{w}^2}{g^2 r^2 e^{ 2\nu}},
\end{split}
\end{equation}
and with $T\indices{^\theta_\theta}$ unchanged. Plugging $T\indices{^r_r}$ into (\ref{r prime squared eq}) and solving for $r'$, we find the desired equation,
\begin{equation} \label{r' eq}
\begin{split}
r^{\prime\,2}
&= e^{2\nu} \left[ 1 - 4\pi \frac{(1 - w^2)^2}{g^2 r^2} \right] 
\\
&\qquad
\times
\left( 1 + 2r\dot{\nu} + \frac{F}{3} r^2 \dot{\nu}^2
- \frac{8\pi \dot{w}^2}{g^2} \right)^{-1}.
\end{split}
\end{equation}
We have succeeded in decoupling the ODEs in (\ref{nu r ODEs}) and finding a field equation for $\nu(r)$.

We are parametrizing the equations with $r$ as the independent variable. To complete the system of equations we must do the same for the equation of motion in (\ref{eom}). Upon converting the $x$ derivatives, we find
\begin{equation} \label{r eom}
\ddot{w} = - \frac{1}{r^{\prime\,2}}  \left[ r'' \dot{w} + \frac{e^{2\nu}}{r^2} w (1-w^2) \right].
\end{equation}
$r''$ can be eliminated using (\ref{r'' convert}) and $r^{\prime\,2}$ can be eliminated using (\ref{r' eq}).

We note that the equations presented in this subsection can be derived directly by assuming a metric of the form
\begin{equation} \label{radial polar metric}
ds^2 = -e^{2\nu(r)} dt^2 + \frac{e^{2\nu(r)}}{b^2(r)} dr^2 + r^2 d\Omega^2,
\end{equation} 
which is parametrized in terms of the metric functions $\nu(r)$ and $b(r)$ and where the areal radius, $r$, is used as the radial coordinate. Comparing this to the metric in (\ref{metric}), we find that $dx = dr/b$ and hence that $b = r'$. Later, we shall reference the $rr$ component of this metric, which is
\begin{equation} \label{radial polar grr}
g_{rr} = \frac{e^{2\nu}}{r^{\prime\,2}}.
\end{equation}

We will make use of the equations presented in this subsection in two places. First, in the next subsection we will derive solutions valid at large $r$, and hence also large $x$. Second, in Sec.~\ref{sec:analytical} we will derive some approximate analytical solutions.


\subsection{Outer boundary conditions}
\label{sec:obc}

In Sec.~\ref{sec:equations}, we constructed the equation of motion in (\ref{eom}) and the field equations in (\ref{nu r ODEs}). These equations are parametrized in terms of the tortoise coordinate, $x$. One might try to use these equations to derive analytical solutions which are valid for $x \rightarrow \infty$. However, such asymptotic solutions are complicated. The solutions are simplified when parametrized in terms of $r$ \cite{Volkov:1998cc}, since they take the simple form 
\begin{equation} \label{asymptotic expansion}
\nu = \nu_0 + \frac{\nu_1}{r} + \frac{\nu_2}{r^2} + \cdots,
\qquad
w = w_0 + \frac{w_1}{r} + \frac{w_2}{r^2} + \cdots.
\end{equation}
In Sec.~\ref{sec:nu(r) w(r)}, we derived the field equation in (\ref{rp nu ODE}) and the equation of motion in (\ref{r eom}) in terms of $r$. We shall use these versions of the equations to derive asymptotic solutions.

We assume that spacetime is asymptotically flat, so that $\nu_0 = 0$. Plugging the asymptotic expansions in (\ref{asymptotic expansion}) into the equations, we find as solutions 
\begin{equation} \label{oBC}
\begin{split}
\nu &= - \frac{M}{r} - \frac{M^2}{r^2} + O(r^{-3})
\\
w &= \pm \left(1 - \frac{w_1}{r} + \frac{3w_1^2 - 6 M w_1}{4r^2} \right) + O(r^{-3}).
\end{split}
\end{equation}
In the bottom equation we can see that $w\rightarrow \pm 1$ as $r\rightarrow \infty$, which is the outer boundary condition used to find classical solutions in Sec.~\ref{sec:classical sols}. The ADM mass $M$, which is the same mass in Eq.~(\ref{ADM M}), enters these solutions by matching the solutions to the asymptotically flat Schwarzschild solution \cite{Volkov:1998cc}. 

Although we do not show this explicitly, the renormalized energy-momentum tensor begins contributing to (\ref{oBC}) at order $r^{-3}$. As a consequence, an important takeaway is that at large $r$, and hence also large tortoise coordinate $x$, the quantum corrected equations are solved by classical solutions. We can therefore use classical solutions as outer boundary values when solving for quantum corrected solutions. 


\subsection{Solution method}

In the previous subsection, we found that we can use classical solutions for our outer boundary values when solving for quantum corrected solutions. We compute classical solutions as described in Sec.~\ref{sec:classical sols}. Classical solutions are computed using the metric in (\ref{classical metric}) and we compute solutions for $\sigma$, $m$, $N$, $w$, $\partial_r w$, and the independent variable $r$. 

Quantum corrected solutions are computed using the metric in (\ref{metric}) and we compute solutions for $\nu$, $\partial_x\nu$, $r$, $\partial_x r$, $w$, $\partial_x w$, and the independent variable $x$. We can convert the variables we use for classical solutions to these variables as follows. The tortoise coordinate, $x$, is related to $r$ by $dx = dr/\sigma N$ and, up to an integration constant, is given by 
\begin{equation}
x = \int \frac{dr}{\sigma(r) N(r)}.
\end{equation} 
The metric field $\nu$ is given by
\begin{equation} \label{nu sigma N}
\nu = \frac{1}{2} \ln \left( \sigma^2 N \right).
\end{equation}
$r$ and $w$ are the same for both sets of variables, while
\begin{equation}
\partial_x w = \sigma N \partial_r w.
\end{equation}
Since we will be solving second order ODEs in (\ref{nu r ODEs}), we need outer boundary values for $\partial_x \nu$ and $\partial_x r$, which are given by
\begin{equation}
\partial_x \nu = \sigma N \partial_r \nu,
\qquad
\partial_x r = \sigma N,
\end{equation}
where
\begin{equation}
\partial_r \nu = 4\pi \frac{r }{N} T\indices{^r_r} +  \frac{m}{r^2 N},
\end{equation}
which follows from (\ref{nu sigma N}) and the classical field equations in (\ref{classical field equations}), with $T\indices{^r_r}$ given in (\ref{classical T}).

We solve for quantum corrected solutions as follows. We compute a classical solution as described in Sec.~\ref{sec:classical sols}. We remind that we label classical solutions by the number of nodes $n$ and horizon radius $\bar{r}_h$. Once we have a classical solution, we choose a value for the coupling constant $g$ and move to the unscaled variables using (\ref{classical scaling}). We then use the formulas presented in this subsection to change variables to those we use for quantum corrected solutions. We use these classical solutions as outer boundary values for the equation of motion in (\ref{eom}) and the field equations in (\ref{nu r ODEs}). We numerically integrate these equations inward, toward smaller $x$. The result is the quantum corrected solution. This solution will include $\partial_x \nu = \nu'$, which can be compared to (\ref{nu' pm}).


\section{Results}
\label{sec:results}

Our computation of quantum corrected solutions begins with classical EYM black hole solutions at the outer boundary. Since classical EYM black holes are identified by $n$, the number of nodes, and $\bar{r}_h$, the horizon radius, we shall also identify quantum corrected solutions by $n$ and $\bar{r}_h$. Additionally, we must specify the coupling constant $g$. We begin by choosing $n=1$ and $g=1$ and presenting solutions for various $\bar{r}_h$ values. Later, we will consider other values for $n$ and $g$.

\begin{figure*} 
\includegraphics[width=7in]{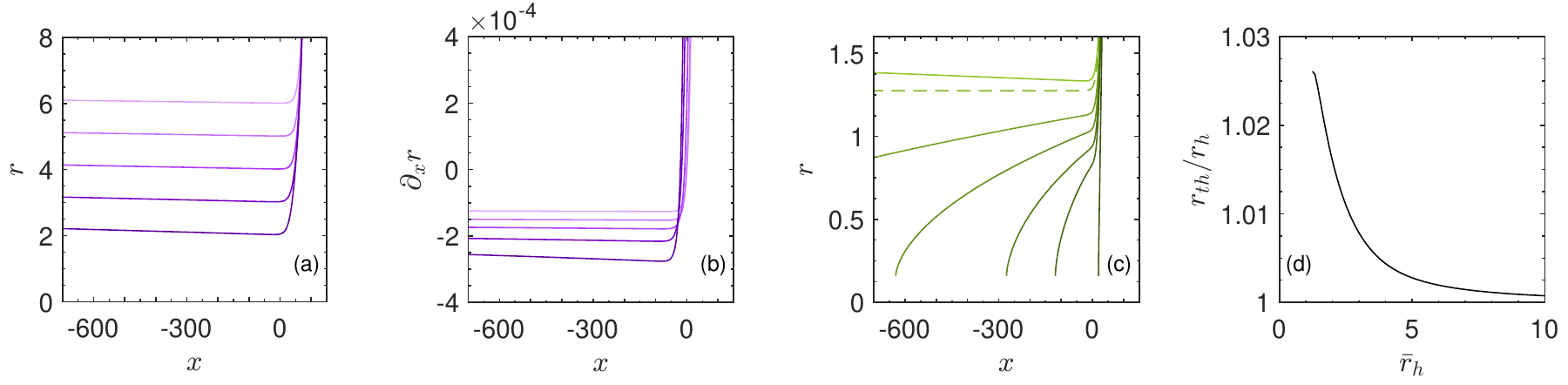} 
\caption{All curves are for quantum corrected EYM black holes with $n=1$ nodes and coupling constant $g=1$. (a) Areal radius $r$ as a function of the tortoise coordinate $x$ for, from top to bottom, $\bar{r}_h = 6,\, 5,\, 4,\, 3,\, 2$. $r$ has a minimum, indicating a wormhole structure. (b) Same curves as (a), but plotting $\partial_x r$. The minimum of $r$ occurs where $\partial_x r$ crosses zero. (c) $r$ as a function of $x$ for, from top to bottom, $\bar{r}_h =  1.3,\, 1.243,\, 1.1,\, 1.0,\, 0.9,\, 0.8,\, 0$. The dashed curve is the critical/transitional solution, below which the wormhole structure no longer occurs. (d) $r_{th}/r_h$, where $r_{th}$ is the wormhole throat radius, i.e.~the value of $r$ at its minimum, as a function of $\bar{r}_h$. The wormhole throat radius is always larger than the classical horizon radius.}
\label{fig:r v x}
\end{figure*}

In Fig.~\ref{fig:r v x}(a), we show the areal radius, $r$, as a function of the tortoise coordinate, $x$, for $\bar{r}_h \geq 2$. We find a wormhole structure, since $r$ has a minimum. The wormhole throat radius, $r_{th}$, is equal to the value of $r$ at its minimum. That this minimum exists can also be established from Fig.~\ref{fig:r v x}(b), since the derivative $\partial_x r$ can be seen crossing zero. Note that $x$ is only defined up to an additive constant. Using this, we have shifted the curves to align them and one should not read into the relative horizontal position of the curves.

The quantum corrected vacuum spacetime, i.e.~quantum corrected Schwarzschild, has been shown to have a wormhole structure \cite{Fabbri:2005zn, Berthiere:2017tms, Ho:2017joh, Ho:2017vgi, Arrechea:2019jgx}. In the quantum corrected vacuum spacetime, $r(x)$ and $\partial_x r(x)$ look very similar to Figs.~\ref{fig:r v x}(a) and \ref{fig:r v x}(b). In our review of classical EYM black holes in Sec.~\ref{sec:classical sols}, we found that YM matter has a minimal effect for larger values of $\bar{r}_h$. Figures \ref{fig:r v x}(a) and \ref{fig:r v x}(b) suggest that this is also the the case for quantum corrected EYM black holes.

YM matter becomes more prominent for quantum corrected solutions as $\bar{r}_h$ decreases, just as it does for classical solutions. In Fig.~\ref{fig:r v x}(c), we again show $r(x)$, but now for $\bar{r}_h < 2$. We find that the minimum of $r$, and hence also the wormhole structure, eventually disappears, making these solutions very different from the vacuum solutions. We have plotted in Fig.~\ref{fig:r v x}(c), as the dashed curve, the transitional solution, which we will call the \textit{critical} solution. The critical solution for $n=1$ and $g = 1$ has $\bar{r}_{hc} \simeq 1.243$. 

For decreasing $x$, quantum corrected solutions with $\bar{r}_h > \bar{r}_{hc}$ have $r \rightarrow \infty$, while quantum corrected solutions with $\bar{r}_h < \bar{r}_{hc}$ have $r \rightarrow 0$.  As $r \rightarrow 0$, solutions approach the divergence in the multiplicative factor in (\ref{upscale}). As this divergence is approached, we lose numerical stability and cannot continue solving for the solution. This is why the curves in Fig.~\ref{fig:r v x}(c) stop before reaching $r = 0$. It is possible to regulate this divergence by choosing a different form for the multiplicative factor, as was done in \cite{Parentani:1994ij, Ayal:1997ab, Arrechea:2019jgx}, but we do not consider that here.

In Fig.~\ref{fig:r v x}(d), we show $r_{th}/r_h$, where $r_{th}$ is the wormhole throat radius, as a function of $\bar{r}_h$ for $\bar{r}_h > \bar{r}_{hc}$. We find that the wormhole throat radius is always larger than the classical horizon. For larger values of $\bar{r}_h$, the wormhole throat radius approaches the classical horizon.

\begin{figure}
\includegraphics[width=3.4in]{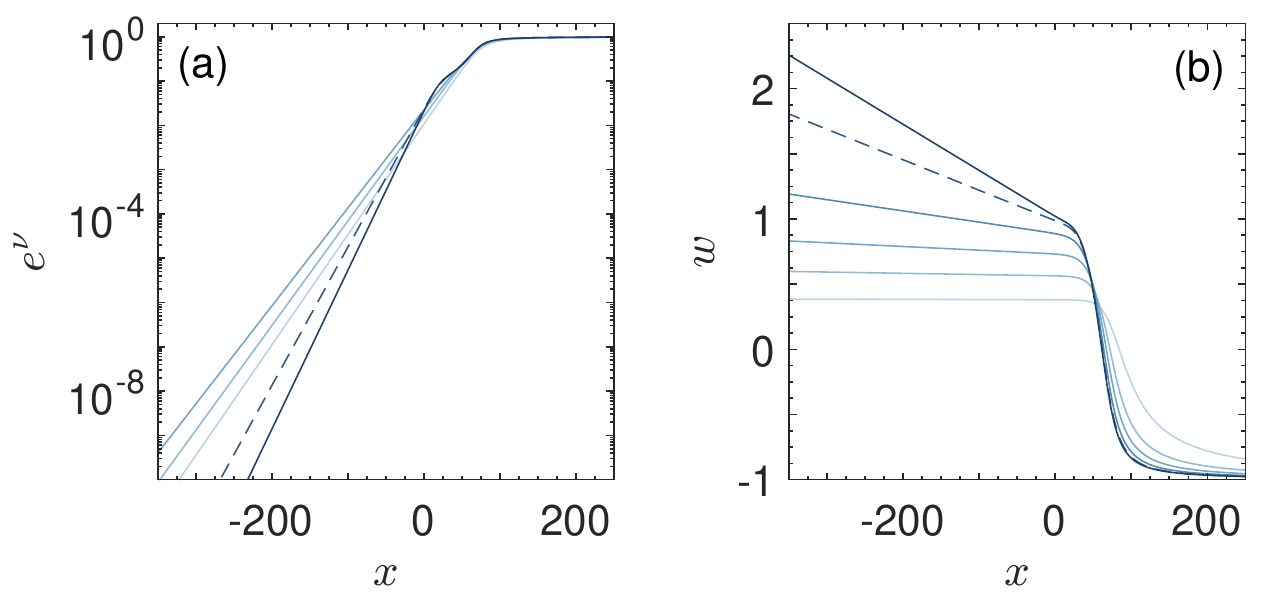} 
\caption{(a) $e^{\nu}$, where $\nu$ is the metric field, as a function of the tortoise coordinate $x$ for, from lightest curve to darkest, $\bar{r}_h = 6,\, 4,\,3,\,1.243,\,1$. (b) Yang-Mills matter field $w$ for, from lightest curve to darkest, $\bar{r}_h = 6,\, 4,\,3,\, 2,\, 1.243,\,1$. All curves have $n =1$ and $g=1$.}
\label{fig:exp(nu) w}
\end{figure}

We show $e^{\nu(x)}$ in Fig.~\ref{fig:exp(nu) w}(a). It is clear that $e^{\nu}$ is nonzero, though it is small, and hence the classical horizon has disappeared. This is unsurprising for $\bar{r}_h > r_{hc}$, since the classical horizon also disappears in the quantum corrected vacuum spacetime \cite{Fabbri:2005zn}. Interestingly, Fig.~\ref{fig:exp(nu) w}(a) shows that the classical horizon also disappears when $\bar{r}_h < \bar{r}_{hc}$. In Sec.~\ref{sec:classical sols}, we reviewed that classical EYM black holes with finite $n$ are nonextremal. Our results are therefore consistent with \cite{Arrechea:2019jgx, Arrechea:2021ldl}, wherein it was argued that including vacuum energy in the form of the renormalized energy-momentum tensor in semiclassical gravity removes classical horizons from nonextremal black holes.

We show the matter field $w(x)$ in Fig.~\ref{fig:exp(nu) w}(b). For larger $\bar{r}_h$, these solutions look similar to their classical counterparts. As $\bar{r}_h$ decreases, these solutions become notably different. Indeed, we find that $w>1$ is possible, which does not occur for classical solutions.

In Sec.~\ref{sec:equations}, we derived Eq.~(\ref{nu' pm}) for $\nu' = \partial_x \nu$, which contains a $\pm$ sign. As explained there, for solutions which use the minus sign, quantum corrections can be thought of as perturbative, while for solutions which use the plus sign, quantum corrections can be thought of as nonperturbative. By comparing solutions with Eq.~(\ref{nu' pm}), we can gain insight into the perturbative and nonperturbative nature of the solution.

\begin{figure}
\includegraphics[width=3in]{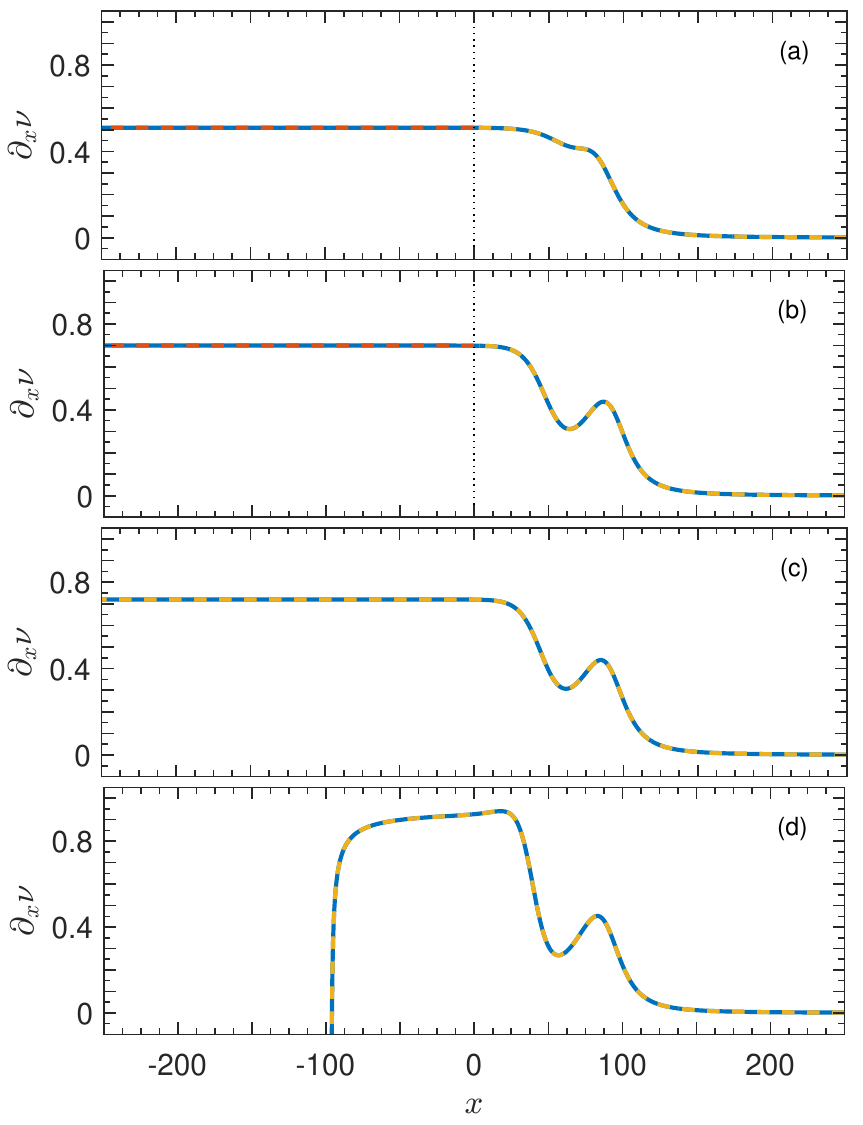} 
\caption{In each plot, the solid blue curve is a solution for a quantum corrected EYM black hole. The dashed yellow curve is Eq.~(\ref{nu' pm}) with the minus sign, which is a perturbative result, and the dashed red curve is Eq.~(\ref{nu' pm}) with the plus sign, which is nonperturbative. The different plots are for (a) $\bar{r}_h = 3$, (b) 1.3, (c) 1.243, (d) 0.8. (a) and (b) are for $\bar{r}_h > \bar{r}_{hc}$ and have a wormhole structure. The curves have been shifted so that the wormhole throat is located at $x = 0$, which is marked by the dotted vertical line. (c) and (d) are for $\bar{r}_h < \bar{r}_{hc}$, with the curve in (c) having $\bar{r}_h$ only slightly smaller than $\bar{r}_{hc}$. All curves have $n = 1$ and $g = 1$.}
\label{fig:nu prime}
\end{figure}

In Fig.~\ref{fig:nu prime}, we show $\nu' = \partial_x\nu$ for four quantum corrected EYM black hole solutions. In each case, the solution is plotted as a solid blue curve. We have then plotted Eq.~(\ref{nu' pm}) with the minus sign as a dashed yellow curve and with the plus sign as a dashed red curve. We have only plotted Eq.~(\ref{nu' pm}), with the relevant sign, for that portion that agrees with the solid blue curve, i.e.~with the solution. Figures \ref{fig:nu prime}(a) and \ref{fig:nu prime}(b) are for $\bar{r}_h > \bar{r}_{hc}$. We have shifted both of these solutions so that the wormhole throat, i.e.~the minimum of $r$, is located at $x = x_{th} = 0$, which is indicated by a vertical dotted line. We can see that the solutions are made up of a mixture of perturbative and nonperturbative solutions. More specifically, for $x>x_{th}$, they are perturbative, while for $x<x_{th}$, they are nonperturbative. The solutions switch continuously from perturbative to nonperturbative precisely at the wormhole throat. A look at Eq.~(\ref{nu' pm}) shows that such a switch can only occur where the square root in (\ref{nu' pm}) vanishes.  We have found that this transition from perturbative to nonperturbative occurring precisely at the location of the wormhole throat occurs in all static solutions containing the wormhole structure that we have studied and we believe it to be a general property. This same phenomenon was seen for quantum corrected vacuum solutions. Interestingly, this changes at the critical solution. For $\bar{r}_h <  \bar{r}_{hc}$, solutions are given entirely by perturbative solutions, as shown in Figs.~\ref{fig:nu prime}(c) and \ref{fig:nu prime}(d). We find that the presence of the wormhole structure requires the nonperturbative result. Since the solutions are asymptotically Schwarzschild, they must also be perturbative. The solution is able to continuously switch from perturbative to nonperturbative and does so precisely at the wormhole throat.

\begin{figure}
\includegraphics[width=3.4in]{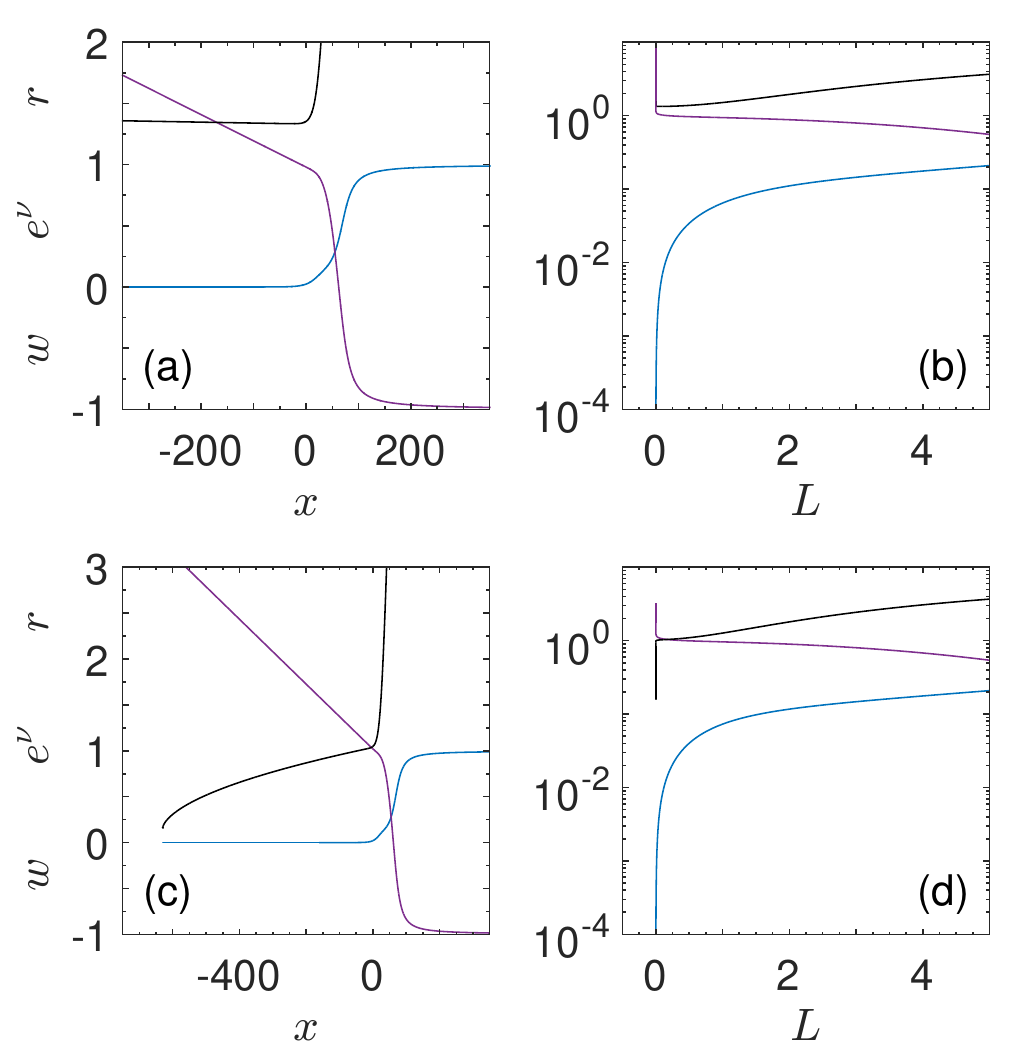} 
\caption{(a) $r$ (black), $e^\nu$ (blue), and $w$ (purple) are shown on the same plot as a function of $x$ for $\bar{r}_h = 1.3 > \bar{r}_{hc}$. (b) The same curves as in (a), but plotted in terms of $L$, the proper distance, showing that a singularity occurs at finite proper distance. The curves have been shifted so that the singularity occurs at $L=0$. (c) and (d) are analogous to (a) and (b), but for $\bar{r}_h = 1 < \bar{r}_{hc}$. In this case, the singularity in the multiplicative factor in (\ref{upscale}) is reached. All curves have $n = 1$ and $g = 1$.}
\label{fig:proper length}
\end{figure}

We show $r$, $e^\nu$, and $w $ in the same plot in Fig.~\ref{fig:proper length}(a) so that we may see how they line up with one another. We stress that $e^{\nu}$, while small, is nonzero, as shown in Fig.~\ref{fig:exp(nu) w}(a). Figure \ref{fig:exp(nu) w}(a) does show $e^{\nu} \rightarrow 0$ asymptotically in $x$ and Fig.~\ref{fig:r v x}(a), for $\bar{r}_h > \bar{r}_{hc}$, shows $r \rightarrow \infty$ asymptotically in $x$. This suggests that there is a singularity as $x \rightarrow -\infty$. In fact, this singularity is located at a finite proper distance. Proper distance is given by
\begin{equation}
L = \int dx\, e^{\nu(x)},
\end{equation}
up to an arbitrary integration constant. We have plotted in Fig.~\ref{fig:proper length}(b) the same curves in Fig.~\ref{fig:proper length}(a), but in terms of $L$. We have shifted these so that the singularity occurs at $L = 0$. Figures \ref{fig:proper length}(c) and \ref{fig:proper length}(d) are another example, but this time for $\bar{r}_h < \bar{r}_{hc}$. In this case, $r\rightarrow 0$ and we approach the divergence in the multiplicative factor (\ref{upscale}).

\begin{figure} 
\includegraphics[width=2.8in]{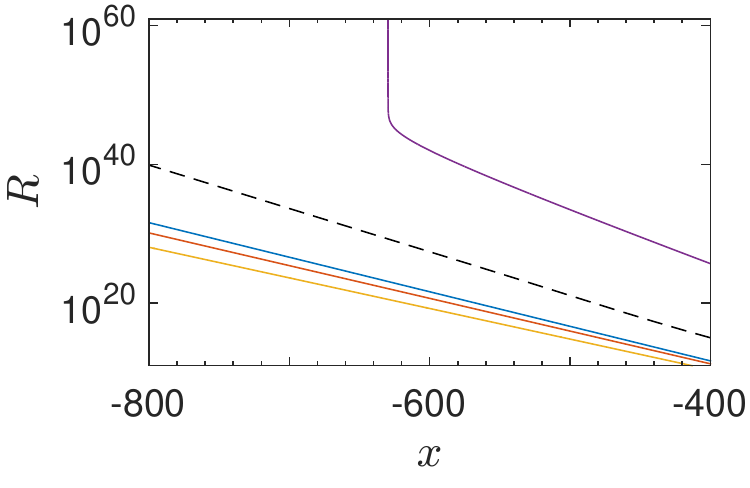} 
\caption{Ricci scalar as a function of the tortoise coordinate for $\bar{r}_h = 6$ (blue), 4 (red), 3 (yellow), 1.243 (dashed black), 1 (purple). The dashed black curve is the critical curve, when $\bar{r} = \bar{r}_{hc}$, and the purple curve is for $\bar{r}_h < \bar{r}_{hc}$. All but the purple curve indicate a curvature singularity as $x\rightarrow -\infty$. The divergence in the purple curve occurs because of the divergence in the multiplicative factor in (\ref{upscale}) as $r\rightarrow 0$. All curves have $n = 1$ and $g = 1$.}
\label{fig:Ricci}
\end{figure}

In Fig.~\ref{fig:Ricci}, we show the Ricci scalar for various solutions. For $\bar{r}_h > \bar{r}_{hc}$, Fig.~\ref{fig:Ricci} indicates a curvature singularity, for which $R\rightarrow \infty$, as $x\rightarrow -\infty$. Such a singularity also occurs in the vacuum solution. The purple curve in Fig.~\ref{fig:Ricci} is a solution for $\bar{r}_h < \bar{r}_{hc}$. As mentioned, for this case $r\rightarrow 0$ and the divergence seen in the purple curve is because we approach the divergence in the multiplicative factor (\ref{upscale}).

\begin{figure} 
\includegraphics[width=2.5in]{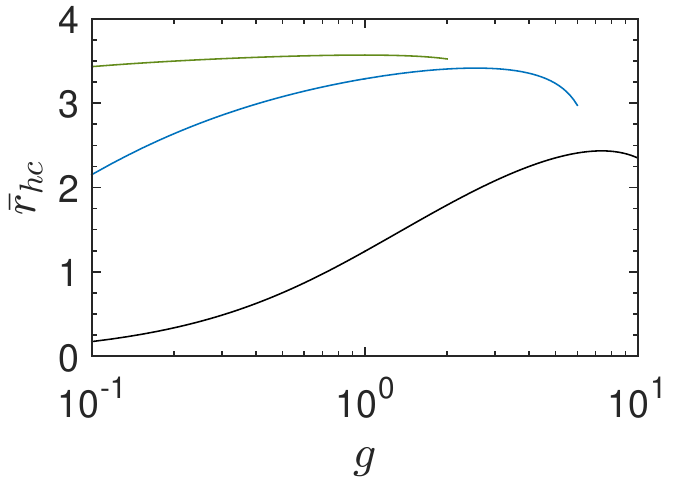} 
\caption{Critical horizon radius, $\bar{r}_{hc}$, for quantum corrected EYM black holes as a function of coupling constant, $g$, for, from bottom to top, $n = 1,\, 2,\, 3$ nodes.}
\label{fig:g}
\end{figure}

We have so far focused only on $n = 1$ and $g=1$. We now consider other values for the number of nodes and the coupling constant. Our interest is in determining if there is new behavior if we change these values. We begin with $n = 1$ and different values for $g$. We have studied quantum corrected solutions for $0.1 < g < 10$ and found them to be qualitatively similar to the $g = 1$ solutions. For example, there exists a critical horizon radius $\bar{r}_{hc}$:~For $\bar{r}_h > \bar{r}_{hc}$, we find a wormhole structure and for $\bar{r}_h < \bar{r}_{hc}$ we do not find a wormhole structure. For all values of the horizon radius, we find that the classical horizon disappears. The critical horizon radius is shown as the bottom curve in Fig.~\ref{fig:g}.

For $n = 2$, the matter field $w$ in the quantum corrected solutions has two nodes, just as it does in the classical solutions. Aside from this, we find qualitatively similar behavior to $n = 1$, as long as $g$ is not too large. We show the critical horizon radius as the middle curve in Fig.~\ref{fig:g} for $0.1 < g< 6$. 

\begin{figure} 
\includegraphics[width=2.5in]{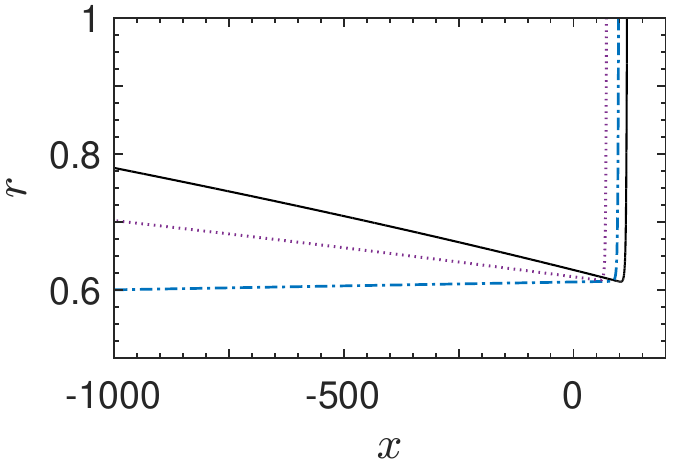} 
\caption{The three curves are for EYM black holes with $n = 2$ nodes and coupling constant $g = 7$. The individual curves have $\bar{r}_{h} = 2$ (solid black), 1 (dash-dotted blue), 0.5 (dotted purple).}
\label{fig:n=2}
\end{figure}

For $g \apprge 6$, we find new behavior. In Fig.~\ref{fig:n=2}, we show quantum corrected solutions for $n=2$ and $g = 7$. The solid black curve has $\bar{r}_h = 2$ and we can see a minimum for $r$. Decreasing $\bar{r}_h$, we find a critical solution, below which the wormhole structure disappears. We show an example of the latter with $\bar{r}_h = 1$ as the blue dash-dotted curve in Fig.~\ref{fig:n=2}. If we continue to lower $\bar{r}_h$, something new happens:~Eventually, the wormhole structure reappears, as shown by the purple dotted curve in Fig.~\ref{fig:n=2}, which has $\bar{r}_h = 0.5$. In all cases, when the wormhole structure is present, solutions are partially perturbative and nonperturbative, similar to Fig.~\ref{fig:nu prime}(a), while if the wormhole structure is absent, solutions are entirely perturbative, similar to Fig.~\ref{fig:nu prime}(c). This behavior does not persist as we increase $g$. Instead, for sufficiently large $g$, we can no longer find solutions that do not have a wormhole structure:~$r$ has a minimum for all values of $\bar{r}_h$.

We have also analyzed $n = 3$, which is qualitatively similar to $n = 2$. $w$ has three nodes, just as it does in classical solutions. We show the critical horizon radius for $0.1 < g < 2$ as the top curve in Fig.~\ref{fig:g}. As $g$ is increased above 2, we find similar behavior to that shown in Fig.~\ref{fig:n=2}. As $g$ is increased further, all solutions have a wormhole structure.

It is interesting that for sufficiently large $g$, we find that the all solutions have a wormhole structure. We have been unable to find a concrete explanation for this and so we present this property of the solutions as an empirical fact.

\begin{figure}
\includegraphics[width=3.4in]{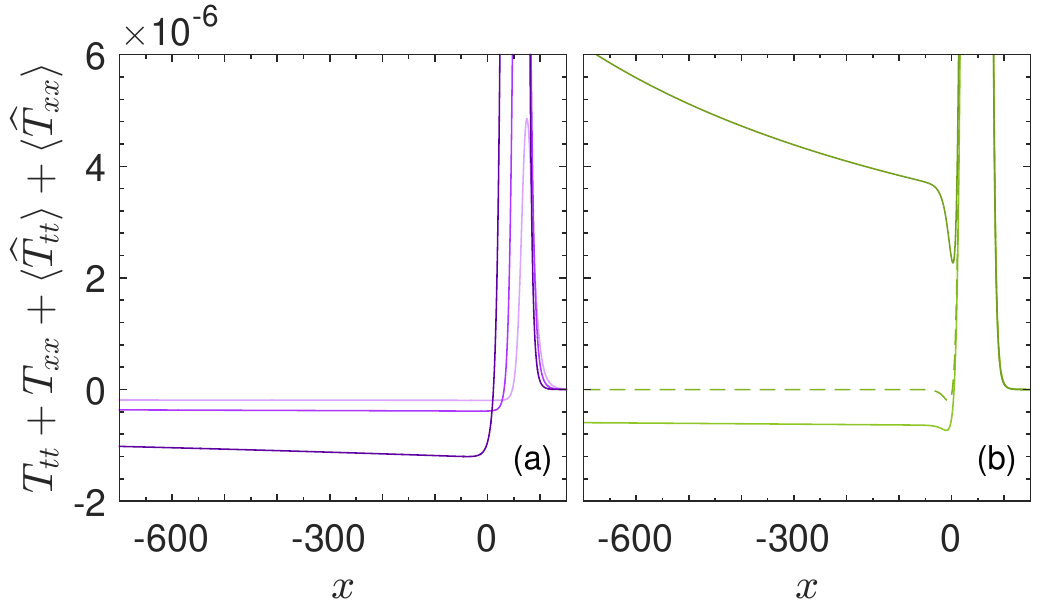} 
\caption{The null energy condition is violated if the curves in (a) and (b) are negative. All curves are for $n = 1$ and $g = 1$. (a) plots solutions for, from top to bottom, $\bar{r}_h = 6, 4, 2$. The corresponding wormhole throats occur at, respectively, $x_{th} = 0.291, -6.857, -20.82$ and the wormholes violate the null energy condition as expected. (b) plots solutions for, from bottom to top, $\bar{r}_h = 1.3, 1.243, 1.1$. For the bottom curve, the wormhole throat occurs at $x_{th} =  -26.06$ and the wormhole violates the null energy condition. The middle dashed curve approximates the critical solution and we can see that this curve is transitional in violating the null energy condition. Finally the top curve, which does not contain a wormhole, does not violate the null energy condition as expected.}
\label{fig:nec}
\end{figure}

For a wormhole to be traversable, it must violate the null energy condition \cite{Morris:1988cz}. Radial null vectors are given by $d^\mu = \lambda(1,\pm 1,0,0)$, where $\lambda$ is an arbitrary constant. A violation of the null energy condition occurs for $(T_{\mu\nu} + \langle \widehat{T}_{\mu\nu} \rangle) d^\mu d^\nu < 0$, which leads to
\begin{equation} \label{nec}
T_{tt} + T_{xx} + \langle \widehat{T}_{tt} \rangle + \langle \widehat{T}_{xx} \rangle < 0.
\end{equation}
Using the classical energy-momentum tensor in (\ref{T}) and the renormalized energy-momentum tensor in (\ref{RSET}), we have that the null energy condition is violated if
\begin{equation} \label{nec2}
\frac{2w^{\prime\,2}}{g^2 r^2}
- \frac{F}{12\pi} \left(\nu^{\prime\,2} - \nu'' \right)
< 0.
\end{equation}
We end this section by plotting the left-hand side of (\ref{nec}) in Fig.~\ref{fig:nec} for some of the solutions shown in Fig.~\ref{fig:r v x}, which are for $n=1$ and $g=1$. Figure \ref{fig:nec}(a) shows solutions with a wormhole. We find that the region of space containing the wormhole throat and beyond violates the null energy condition, as expected. Figure \ref{fig:nec}(b) shows that solutions without a wormhole do not violate the null energy condition, also as expected. It is the renormalized energy-momentum tensor that allows the left-hand side of (\ref{nec}) to become negative. We note that the spikes, on the positive side of $x = 0$, in Fig.~\ref{fig:nec}, occur because $w$ has a large derivative, as can be seen in Fig.~\ref{fig:exp(nu) w}(b), which contributes the classical energy-momentum tensor in (\ref{nec2}).


\section{Analytical approximations}
\label{sec:analytical}


\subsection{Wormhole throat}

In this subsection, we derive an approximate analytical formula for the geometry near the wormhole throat. This is the geometry that replaces the classical horizon in the quantum corrected solution. The wormhole throat is located at the minimum of $r$. Using $x$ as the independent variable, the wormhole throat is located where $r' = \partial_x r = 0$. It is possible to use $\nu$ as the independent variable. Then the wormhole throat is located where $\partial_\nu r = 0$. Inverting this, the wormhole throat is located where $\dot{\nu} = \partial_r \nu \rightarrow \infty$. Near the wormhole throat, then, both $\dot{\nu}$ and $1/r'$ are large.

We can find an approximate analytical formula that is valid near the wormhole throat by dropping subdominant terms in (\ref{rp nu ODE}), which gives
\begin{align} \label{ddot nu approx}
\ddot{\nu} \left(1 - \frac{F}{3} \right)
&\simeq 
4\pi r e^{2\nu}\frac{\dot{\nu}}{r^{\prime\,2}}  \left[ \frac{2 w^{\prime\,2}}{g^2 r^2 e^{ 2\nu}} + \frac{2 F}{3} \frac{(1 - w^2)^2}{2g^2 r^4}\right]
\notag
\\
&\qquad
- \frac{F r}{3} \dot{\nu}^3 \left[1 + \frac{1}{3}  \left(F + \frac{r F'}{2r'} \right) \right],
\end{align}
where we plugged in for the energy-momentum tensor components using (\ref{T}). Note that $w' = r' \dot{w}$. $r'$ is small while $\dot{w} = (\partial r/ \partial w)^{-1}$ is large near the wormhole throat. As such, we cannot assume  $w'$ is small or large. We could replace $r^{\prime \,2}$ in (\ref{ddot nu approx}) using (\ref{r' eq}), except that the derivation of (\ref{r' eq}) made use of $w' = r' \dot{w}$. Rederiving (\ref{r' eq}) by returning to the $rr$ component of the field equations in (\ref{rr field equation}), but this time retaining $w'$ and dropping subdominant terms, we find
\begin{equation} \label{grr eq}
\frac{e^{2\nu}}{r^{\prime\,2}} \simeq \frac{F}{3} r^2  \dot{\nu}^2
\left\{ 1 + 8\pi \left[\frac{w^{\prime\,2}}{g^2 e^{ 2\nu}}  -\frac{(1 - w^2)^2}{2g^2 r^2} \right] \right\}^{-1}.
\end{equation}
Inserting this into (\ref{ddot nu approx}), we arrive at
\begin{equation} \label{dot nu approx}
\frac{d\dot{\nu}}{\dot{\nu}^3}
\simeq - dr \frac{Fr/3}{1- F / 3} 
\left[1 + \frac{1}{3}  \left(F + \frac{r F'}{2r'} \right) 
- W
\right],
\end{equation}
where
\begin{equation} \label{Wdef}
\begin{split}
W &\equiv 4\pi r^2
\left[ \frac{2 w^{\prime\,2}}{g^2 r^2 e^{ 2\nu}} + \frac{F}{3} \frac{(1 - w^2)^2}{g^2 r^4}\right]
\\
&\qquad \times
\left\{ 1 + 8\pi  \left[\frac{w^{\prime\,2}}{g^2 e^{ 2\nu}}  -\frac{(1 - w^2)^2}{2g^2 r^2} \right] \right\}^{-1}.
\end{split}
\end{equation}

\begin{figure} 
\includegraphics[width=2.75in]{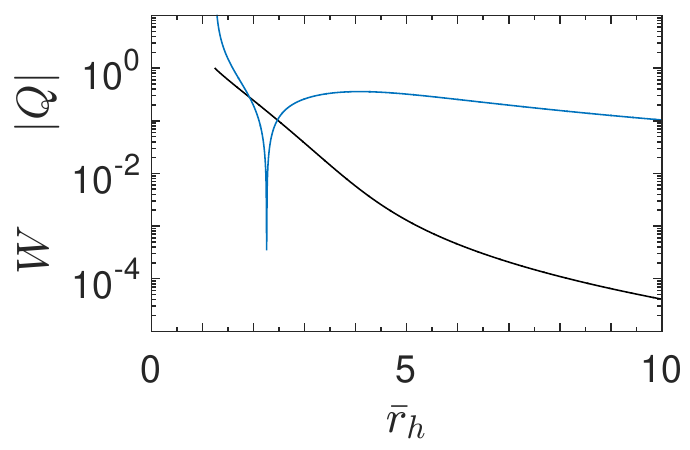} 
\caption{$W$ (black curve), defined in (\ref{Wdef}), and $|Q|$ (blue curve), defined in (\ref{Qdef}) as a function of $\bar{r}_h$ with $n = 1$ and $g = 1$.}
\label{fig:WQ}
\end{figure}

We have been unable to find a useful analytical approximation for $W$ that is valid near the wormhole throat. We therefore resort to numerically evaluating $W$. We show $W$, evaluated at the wormhole throat, for $n=1$ and $g=1$ as the lower black curve in Fig.~\ref{fig:WQ}. We find that $W \ll 1$ unless the quantum corrected solution is approaching the critical solution. When not considering solutions near the critical solution we can safely neglect $W$ in (\ref{dot nu approx}), which then reduces to 
\begin{equation} \label{approx1}
\frac{d\dot{\nu}}{\dot{\nu}^3}
\simeq - dr \frac{Fr/3}{1- F / 3} 
\left[1 + \frac{1}{3}  \left(F + \frac{r F'}{2r'} \right) 
\right].
\end{equation}
Notice that all reference to matter has vanished, making this the quantum corrected vacuum formula. Using (\ref{upscale}) for $F$, Eq.~(\ref{approx1}) is integrable. We find that the geometry near the wormhole throat is described by
\begin{equation} \label{nu wormhole}
\nu \simeq \sqrt{ \frac{2(12\pi r_{th}^2 - 1)}{r_{th}}}
\sqrt{r - r_{th}} + \nu_{th},
\end{equation}
where we expanded about the wormhole throat to obtain this result and where $r_{th}$ and $\nu_{th}$ are the values of $r$ and $\nu$ at the wormhole throat.

Equation (\ref{nu wormhole}) is the same result found for the quantum corrected vacuum spacetime \cite{Fabbri:2005zn, Ho:2017joh, Arrechea:2019jgx}. However, it is not the case that the full geometry near the wormhole is well approximated by the vacuum spacetime, even far from the critical solution. This can be seen by considering the metric in (\ref{radial polar metric}) and in particular the $rr$ component of this metric given by (\ref{radial polar grr}). Using (\ref{grr eq}), we have for the metric in (\ref{radial polar metric}) that
\begin{equation} \label{Qdef}
\begin{split}
g_{rr} &\simeq \frac{F}{3} r^2  \dot{\nu}^2
( 1 + Q )^{-1}
\\
Q &\equiv 8\pi \left[\frac{w^{\prime\,2}}{g^2 e^{ 2\nu}}  -\frac{(1 - w^2)^2}{2g^2 r^2} \right].
\end{split}
\end{equation}
We show $|Q|$, evaluated at the wormhole throat, for $n=1$ and $g=1$ as the upper blue curve in Fig.~\ref{fig:WQ}. We find that $Q$ is not much smaller than 1. As a consequence, we cannot ignore the matter dependence in $g_{rr}$ and $g_{rr}$ is not well approximated by the vacuum formula. This also means that we cannot find an analytical approximation for proper length.


\subsection{Critical solution}

We can derive an approximate analytical formula for the critical solution. This formula is derived from the $rr$ component of the field equations in (\ref{rr field equation}), where $T\indices{^r_r}$ is given in (\ref{T}). At the critical solution, $r' = 0$. We have seen numerically that throughout the region where $r' = 0$, $e^{2\nu}$ is extremely small. Dropping both $r'$ and $e^{2\nu}$ we find 
\begin{equation}
\nu' \simeq 12\pi \sqrt{\frac{2}{3}} \frac{w'}{g},
\end{equation}
where we plugged in (\ref{upscale}) for the multiplicative factor $F$. This is straightforward to integrate,
\begin{equation} \label{critical nu}
\nu \simeq 12\pi \sqrt{\frac{2}{3}} \frac{w}{g} + C,
\end{equation}
where $C$ is an integration constant. As mentioned, this equation is valid for the critical solution in the region where $r$ is constant.

Using that $e^{2\nu}$ is small in the equation of motion in (\ref{eom}), we have $w'' \simeq 0$ and that $w$ is linear in $x$. Using this in (\ref{critical nu}), we have that $\nu$ is also linear in $x$.


\section{Conclusion}
\label{sec:conclusion}

Einstein-Yang-Mills black holes are classical black hole solutions in static spherically symmetric general relativity \cite{Volkov:1989fi, Bizon:1990sr, Kuenzle:1990is}. They were the first hairy black hole solutions discovered \cite{Volkov:2016ehx} and are constructed using $SU(2)$ Yang-Mills matter within the magnetic ansatz. In this work, we generalized these solutions by including a renormalized energy-momentum tensor in the Polyakov approximation within the framework of semiclassical gravity.

We found quantum corrected EYM solutions by solving the field equations and the equation of motion self-consistently. Our solutions bear similarities to the quantum corrected vacuum solution \cite{Fabbri:2005zn, Berthiere:2017tms, Ho:2017joh, Ho:2017vgi, Arrechea:2019jgx, Arrechea:2021ldl}. In both cases, the classical horizon is replaced with a wormhole structure. However, we also found solutions for quantum corrected EYM black holes in which the classical horizon still disappears, but there is no wormhole structure. Our quantum corrected solutions are identified by the number of radial nodes, the classical horizon radius, and the coupling constant. We presented a broad analysis of the quantum corrected solutions for various values of these parameters.

It has been argued, at least for static spherically symmetric solutions, that the inclusion of a renormalized energy-momentum tensor for nonextremal classical black holes in semiclassical gravity will remove the classical horizon \cite{Arrechea:2019jgx, Arrechea:2021ldl}. This was found to be the case in vacuum and in electrovacuum, i.e.~for quantum corrected Schwarzschild and Reissner–Nordstr\"om black holes. Our results indicate that this continues to be the case in the presence of black hole hair. That this occurs in a wide range of classical black hole solutions indicates that the disappearance of the black hole horizon in semiclassical gravity is a robust phenomenon.

An important question is whether the static solutions are stable with respect to time dependent perturbations. Studies so far have restricted themselves to static spacetimes. One method for determining stability is to use a static solution for the initial conditions, along with a small perturbation, and to allow the system to evolve in time. If the system holds its configuration, the static solution is stable. If the solution evolves to some new configuration, the static solution is unstable. If the static solutions are stable, then it is possible that the solutions could occur naturally. On the other hand, if the static solutions are unstable, then they could not occur naturally, at least not as a final state. A determination of stability therefore sheds light on whether these solutions are of practical or just academic interest. We are currently pursuing this question of stability and will report results elsewhere.




%

\end{document}